\documentclass[10pt]{amsart}

\usepackage{amsmath}
\usepackage{amssymb}
\usepackage{graphicx}
\usepackage{subfigure}
\usepackage{cite}

\title{Formation dynamics of black- and white-hole horizons in an analogue gravity model}
\author{Manuele Tettamanti $^{1,2}$ and Alberto Parola $^{3}$}
\address{%
$^{1}$ \quad Dipartimento di Fisica ``Giuseppe Occhialini", Universit\`a di Milano-Bicocca, Piazza della Scienza 3, 20126 Milano, Italy \\
$^{2}$ \quad INFN — Sezione di Milano-Bicocca, Piazza della Scienza 3, 20126 Milano, Italy \\
$^{3}$ \quad Dipartimento di Scienza e Alta Tecnologia, Universit\`a degli Studi dell’Insubria, Via Valleggio 11, 22100 Como, Italy}

\begin{document}

\begin{abstract}
{We investigate the formation dynamics of sonic horizons in a Bose gas confined in a (quasi) one-dimensional trap. This system is one of the most promising realizations of the analogue gravity paradigm and has already been successfully studied experimentally. Taking advantage of the exact solution of the one-dimensional, hard-core, Bose model (Tonks-Girardeau gas) we show that, by switching on a step potential, either a sonic (black-hole-like) horizon or a black/white hole pair may form, according to the initial velocity of the fluid. Our simulations never suggest the formation of an isolated white-hole horizon, although a stable stationary solution of the dynamical equations with those properties is analytically found. Moreover, we show that the semiclassical dynamics, based on the Gross-Pitaevskii equation, conforms to the exact solution only in the case of fully subsonic flows while a stationary solution exhibiting a supersonic transition is never reached dynamically.}
\end{abstract}

\maketitle

\section{Introduction}

Analogue models of gravity are useful tools bridging gravitation to other branches of physics and they have been intensively investigated since their proposal \cite{unruh} in order to study effects whose experimental achievement is hardly possible in the cosmological context. In the past years a great scientific effort was dedicated to the observation and characterization of the Hawking effect \cite{hawking1,hawking2} but several different phenomena can be investigated 
within the analogue gravity paradigm \cite{liberati}. In particular, the physics of sonic horizons \cite{barcelo} has been vastly studied, both from a theoretical and an experimental point of view and analogue black-hole and white-hole horizons have been experimentally achieved in optics \cite{Philb,sergio,drori}, in Bose-Einstein condensates (BECs) \cite{stein2010,stein14,stein2016,stein18,stein19}, in water \cite{rous08,jannes,silk,rous15,rous16}, in superfluid Helium \cite{he3} and a in few other setups. 
In the context of BECs, a few recent experiments \cite{stein14,stein2016,stein18,stein19} successfully recreated sonic horizons in order to probe the elusive Hawking radiation but the results are still under debate in the scientific community \cite{tettamanti,ted1,ted2,stein-numerical,ulf_jeff,jeff_response,epl,prd,isoard}. Nonetheless, these experiments represent a remarkable achievement and they have stimulated a variety of theoretical studies on sonic black holes in 1D BECs. One feature, though, which has not been deeply investigated in these kind of configurations is the formation dynamics of an analogue black/white hole; this may be due to the fact that analogue models can hope to reproduce only the kinematical aspects of the gravitational phenomena, as the analogy breaks down for the dynamical properties. For this reason, often theoretical studies start from an ``eternal'' (analogue) black-hole, i.e., from a configuration which already shows a sonic horizon, neglecting the question of how it may have formed.\\

In this work we will study the dynamics of formation of a sonic horizon for a simple, realistic model of BEC which mimics existing experimental configurations \cite{stein2010,stein14,stein2016,stein18,stein19}. In particular, we will exploit a recently-proposed exact model \cite{epl,prd} in order to study the dynamics of the analogue of a gravitational collapse in the absence of external confining traps and to investigate whether the system only forms black-hole-like horizons or if a dynamical process could, in principle, lead to a formation of a white-hole horizon too. Furthermore, we will compare the results with those obtained with a semiclassical approach - commonly adopted in the theoretical interpretations of the experiments - in order to highlight possible inadequacies of the semiclassical paradigm. This will give insights on the formation process of analogue horizons and, interestingly, will also lead to challenge the ``eternal'' black-hole configuration as a tool to study the Hawking effect.

\section{Tonks-Girardeau gas}

We consider a one-dimensional Bose fluid of hard-core, point-like particles in an external potential $V(x)$. The model hamiltonian is written, in first quantization, as

\begin{equation}
H=\sum_{i=1}^N\left [ -\frac{\hbar^2}{2m}\frac{d^2}{dx_i^2} +V(x_i)\right ] + c\sum_{i<j}\delta(x_i-x_j) \, ,
\label{ham}
\end{equation}

where the hard-core constraint corresponds to the strong interaction limit $c\to\infty$. In the absence of external potential $V(x)=0$, eigenvalues and eigenfunctions of this hamiltonian have been found for any value of $c$ by Lieb and Liniger \cite{ll} using Bethe Ansatz techniques. In the hard-core limit, the Lieb-Liniger solution can be more easily obtained by performing a Jordan-Wigner transformation which maps the hard core Bose fluid into a non-interacting, spinless, fermion model \cite{gira}. A simple argument allows to understand this exact result: in one dimension the spatial ordering of hard-core bosons cannot be modified by a local hamiltonian. Therefore, the different phases acquired by bosons and fermions upon particle exchange do not come into play and the matrix element of the hamiltonian between two arbitrary bosonic configurations coincides with the corresponding matrix element of the free spinless Fermi gas. Interestingly, this result remains valid also in the presence of an arbitrary external potential, but clearly the argument fails in more than one dimension, where a natural ordering cannot be defined. Due to this mapping, the full spectrum of the hamiltonian (\ref{ham}) and the many-particles dynamical density correlations coincide with those of a non-interacting Fermi fluid in the same external potential $V(x)$. Note that, instead, the momentum distribution of the Bose gas differs from that of the corresponding Fermi system due to the non-locality of the Bose-Fermi mapping defined by the Jordan-Wigner transformation.

Exactly solvable models represent a unique tool as a test ground for approximate theories, which are the only possible option in most cases. Among the approximate descriptions of a weakly interacting Bose fluid, the celebrated Gross-Pitaevskii equation (GPE) \cite{GPE} stands out as a simple and accurate method to portray the dynamics of the condensate wavefunction. The inclusion of a cubic non-linearity in the Schr\"odinger equation suffices to account for weak interparticle interactions, providing a satisfactory description of many experiments in cold atom physics \cite{GPE}. The Tonks-Girardeau (TG) gas is a strongly interacting Bose fluid and therefore the usual semiclassical approaches for the approximate description of the dynamics are not expected to provide accurate results. However, it is known that a simple generalization of the Gross-Pitaevskii equation does indeed faithfully reproduce the physics of a Tonks-Girardeau gas \cite{Stringari,Salasnich}, at least in homogeneous systems. Such a ``strong coupling limit" of the usual (cubic) GPE displays a quintic non-linearity with a specific value of the coupling constant $g=\frac{\hbar^2\pi^2}{2m}$:

\begin{equation}
i\,\hbar  \frac{\partial\psi(x,t)}{\partial t} = -\frac{\hbar^2}{2m}\,\frac{\partial^2\psi(x,t)}{\partial x^2} + g \,|\psi(x,t)|^4\,\psi(x,t) + V(x)\,\psi(x,t) \, .
\label{gpe5}
\end{equation}

In the Gross-Pitaevskii approximation, the global properties of the flowing Bose gas are identified with those derived from the single particle wavefunction $\psi(x,t)$. In particular, in the homogeneous case ($V(x)=0$), the phonon excitation spectrum derived from Eq. (\ref{gpe5}) coincides with the exact one, defined by a sound velocity $c$, related to the fluid density $n$ by the same expression valid for a one dimensional Fermi gas: $c=\pi \frac{\hbar n}{m}$. The dynamics described by the Gross-Pitaevskii equation will be compared with the exact results in order to verify the accuracy of the widely adopted semiclassical approximation in the context of analogue gravity experiments in BECs. 

\section{The model}

The Tonks-Girardeau gas model has been recently studied in the framework of analogue gravity \cite{epl,prd} because it allows to describe, from a microscopic point of view, the dynamics of formation of a sonic horizon and to investigate the physical conditions leading to the expected Hawking-like phonon flux emerging from the horizon. The analytical and numerical studies of Ref. \cite{epl,prd} proved that a precise correspondence between the physics of the analogue model and that of the Hawking effect can be obtained only when the TG gas flows against an extremely smooth potential barrier. The rationale for this result is that the identification of the quasiparticles (phonons) as a free quantum field is possible only for very low excitation energies and (almost) homogeneous states. Instead, sharp potentials, like those usually adopted in the experiments, give rise to a steady state characterized by a rapidly varying density profile, whose elementary excitations cannot be expressed as a free field with a locally varying sound velocity, as requested by the analogy to the gravitational problem. Remarkably, a sharp external potential does not seem to suppress the Hawking effect: in the steady state a phonon flux may indeed be present but it is not described by a thermal distribution and therefore it cannot be associated to a well-defined Hawking temperature. We stress that these results have been obtained \cite{epl,prd} for the Tonks-Girardeau gas in one dimension, where the occurrence of thermalization after a quantum quench is hampered by the integrability of the model \cite{thermal}. On the other hand, at variance with the case of a potential barrier, an extremely smooth potential step does not lead to the formation of a sonic horizon, making such a configuration unsuitable for the investigation of the Hawking effect \cite{epl,prd}. Finally, it has been thoroughly discussed in \cite{epl,prd} that, even though the hard core limit may appear as a strong assumption which is hardly ever achieved experimentally \cite{kino,bloch}, the conclusions drawn from that model can be generalized also for the case of weakly interacting gases. This, as we will show, can be said for the results derived in this paper too.\\

In this work we will examine the formation dynamics of black- and white-hole horizons when a flowing TG gas is perturbed by the sudden switching-on of a sharp step potential. In the experiments, this procedure has been adopted in quasi one-dimensional BECs in order to create either a sonic black-hole horizon \cite{stein2016,stein18,stein19} or a black/white hole pair in order to trigger the so-called ``black-hole laser'' effect \cite{stein14}. As the analogue gravity paradigm shows, the first is obtained when the fluid passes from subsonic to supersonic (in the direction of the flow) while a white-hole horizon is the exact opposite, i.e. when the fluid passes from a supersonic region to a subsonic one. Indeed, in the first case phonons cannot exit the supersonic region (the analogue of the interior of a black hole) while in the second they are forced to do so (the analogue of a white hole indeed). According to the experimental configurations and to our previous theoretical studies \cite{epl,prd}, we choose a left-moving gas flowing over a "waterfall" potential as a preferred setup but we will generalize our investigation to also describe the opposite flow.

Thus, we start from a flowing TG gas in a stationary state which is described, via the Bose-Fermi mapping, by a Slater determinant of plane waves $\frac{1}{\sqrt{2\pi}}e^{ipx}$; the wavevectors $p$ belong to the interval $-k_F-k_0 \le p \le k_F-k_0$, where $k_F>0$ is the Fermi momentum in the co-moving reference frame and $k_0$ is the momentum shift due to the fluid flow; these two parameters, characterizing the initial state of the system, are related to the uniform particle density $n = \frac{k_F}{\pi}$, sound speed $c=\frac{\hbar k_F}{m}$ and fluid velocity $v=-\frac{\hbar k_0}{m}$. We therefore adopt the convention that a positive value of $k_0$ corresponds to a left-moving flow. Then, at $t=0$, we suddenly turn on the step potential 

\begin{equation}
V(x) =
\begin{cases}
0 & \text{for} \quad x < 0 \\
V_0 & \text{for} \quad x > 0
\end{cases} \, ;
\end{equation}

where $V_0$ is the potential height which is conveniently parametrized\footnote{In the following, we will use the wavector $Q$ in order to make lengths adimensional.} through the wavevector $Q$: $V_0 = \frac{\hbar^2Q^2}{2m}$. Immediately after the quench the fluid is not in a stationary state any more and its many particle wavefunction evolves in time. Adopting the fermionic representation, at any time $t>0$ the wavefunction is given by the Slater determinant of the time evolution of the initial plane waves: 

\begin{equation}\label{evol}
\psi_p(x,t) = \int_{-\infty}^{\infty} dk \, c_p(k) \,\phi_k(x)\,e^{-i\omega_k t} \, ,
\end{equation}

\begin{equation}\label{evol1}
c_p(k) = \int_{-\infty}^{\infty} \frac{dx}{\sqrt{2\pi}} \, \phi^*_k(x)\,e^{ipx}\,e^{-\eta |x|} \, ,
\end{equation}

where $\phi_k(x)$ are the exact single particle eigenfunctions in the presence of the external potential, $\epsilon_k = \hbar\omega_k$ the corresponding energy, and $\eta\to 0^+$ is the usual convergence factor. For a step potential $\epsilon_k = \frac{\hbar^2k^2}{2m}$ and the eigenfunctions $\phi_k(x)$ are easily obtained by elementary methods: 

\begin{itemize}
\item for $0<k \le Q$
\begin{equation}
\phi_{k}(x) =
\begin{cases}
\frac{1}{\sqrt{2\pi}}\,\left [ e^{ikx}+R_k\,e^{-ikx}\right ] & \text{for} \quad x < 0 \\
\frac{1}{\sqrt{2\pi}}\,T_k\,e^{-\lambda_kx}& \text{for} \quad x > 0
\end{cases} \, ;
\end{equation}
\item for $k>Q$  
\begin{equation}
\phi_{k}(x) =\begin{cases}
\frac{1}{\sqrt{2\pi}}\,\left [ e^{ikx}+R_k\,e^{-ikx}\right ] & \text{for} \quad x < 0 \\
\frac{1}{\sqrt{2\pi}}\,T_k\,e^{iqx}& \text{for} \quad x > 0
\end{cases} \, ;
\end{equation}
\item for $k<-Q$ 
\begin{equation}
\phi_{k}(x) =\begin{cases}
\frac{1}{\sqrt{2\pi}}\,\sqrt{\frac{|k|}{q}}\,T_k\,e^{ikx}
& \text{for} \quad x < 0 \\
\frac{1}{\sqrt{2\pi}}\,\sqrt{\frac{|k|}{q}}\,\left [ e^{-iqx}+R_k\,e^{iqx}\right ] 
& \text{for} \quad x > 0
\end{cases} \, ;
\label{left}
\end{equation}
\end{itemize}

while no eigenfunction exists for $-Q<k<0$. Here $\lambda_k = \sqrt{Q^2-k^2}$ and $q=\sqrt{k^2-Q^2}$. The explicit expressions of the reflection and transmission coefficients are reported in \cite{prd}.

\section{Steady states}\label{station}

While the full dynamics after the quench requires a numerical analysis starting from Eqs. (\ref{evol}-\ref{evol1}), the long time behavior can be analytically evaluated. As shown in Ref. \cite{prd}, at any given fixed position $x$, the local physical properties of the model approach those given by the stationary state built upon the wavefunctions $\psi_p(x,t)$ given by: 

\begin{equation}
\psi_p(x,t) = \begin{cases}
\phi_p(x)\,e^{-i\omega_p t} & \text{for} \quad p > 0 \\
\sqrt{\frac{|p|}{k}}\,\phi_{-k}(x)\, e^{-i\omega_k t} & \text{for} \quad p < 0 
\end{cases} \, ,
\label{asy}
\end{equation}

with $k=\sqrt{p^2+Q^2}$. As a consequence, for each choice of the two parameters $(k_F,k_0)$ a steady state exists, characterized by the number density profile

\begin{equation}
n(x) = \int_{-k_F-k_0}^{k_F-k_0} dp\, |\psi_p(x)|^2 
\label{rho}
\end{equation}

and by the mass current

\begin{equation}
j = i\frac{\hbar}{2}\,\int_{-k_F-k_0}^{k_F-k_0} dp \left [ \frac{d\psi_p^*(x)}{dx}\,\psi_p(x) -  \frac{d\psi_p(x)}{dx}\,\psi_p^*(x)\right ] \, ,
\label{gei}
\end{equation}

which is constant in space and time. By direct substitution of the scattering eigenfunctions in Eq. (\ref{asy}) we get the explicit results:

\begin{equation}\label{analytic1}
j =-\frac{2\hbar}{\pi} \left [\int_{-\infty}^{0} dp\,f(p)\,\frac{kp^2}{(k-p)^2} -\int_{Q}^{\infty}dp\,f(p)\,\frac{qp^2}{(p+q)^2} \right ] \, ,
\end{equation}

\begin{eqnarray}\label{analytic2}
n(x)& = & \frac{1}{\pi} \left [2\,\int_{-\infty}^{0} dp\,f(p)\,\frac{p^2} {(k-p)^2} +\int_{0}^{Q} dp\, f(p)\left [ 1+ \cos \left (2px + 2\,\arccos{\frac{p}{Q}}\right ) \right ] \right. +  \nonumber \\	
& &  \qquad \qquad \left. +\int_{Q}^{\infty} dp\,f(p)\,\frac{p^2+q^2+Q^2\,\cos 2px)}{(q+p)^2} \right ] 
\end{eqnarray}

for $x < 0$ and

\begin{eqnarray}\label{analytic3}
n(x)&=&\frac{1}{\pi} \left [\int_{-\infty}^{0} dp\,f(p)\,\left [ \frac{p^2+k^2} {(k-p)^2} - \frac{k+p} {k-p}\,\cos 2px \right ] +2\,Q^{-2}\int_{0}^{Q} dp\, f(p)\,p^2\,e^{-2\sqrt{Q^2-p^2}\,x} \right. + \nonumber \\  & & \qquad \qquad \qquad \left. + \int_{Q}^{\infty} dp\,f(p)\,\frac{2p^2}{(p+q)^2}\right ] 
\end{eqnarray}

for $x > 0$, where $q=\sqrt{p^2-Q^2}$ and $k=\sqrt{p^2+Q^2}$. The occupation number $f(p)$ identifies the region in momentum space occupied by fermions: $f(p)=1$ for $-k_F-k_0 \le p \le k_F-k_0$ and $f(p)=0$ elsewhere. These expressions allow to evaluate several important quantities in the framework of the analogue gravity: the local fluid and sound velocity, $v(x)$ and $c(x)$ respectively, together with their ratio, i.e. the Mach number:

\begin{eqnarray}
v(x) &=& \frac{j}{m\,n(x)} \, , \\
c(x) &=& \pi\frac{\hbar \,n(x)}{m} \, , \\
\frac{|v(x)|}{c(x)}&=& \frac{|j|}{\pi \hbar\,n(x)^2} \, .
\end{eqnarray}

The analytical expressions show that in the limiting cases $k_0 \geq k_F$ the density profile is rigorously flat in the region $x<0$, leading to a uniform fluid velocity. The same can be said for the region $x>0$ when $-k_0 \geq Q+k_F$. At first, though, we will consider only the cases of $|k_0| \le k_F$ as the condition ensures that the flow is initially subsonic.

\begin{figure} [ht!]
\begin{center}
\includegraphics[width=8cm]{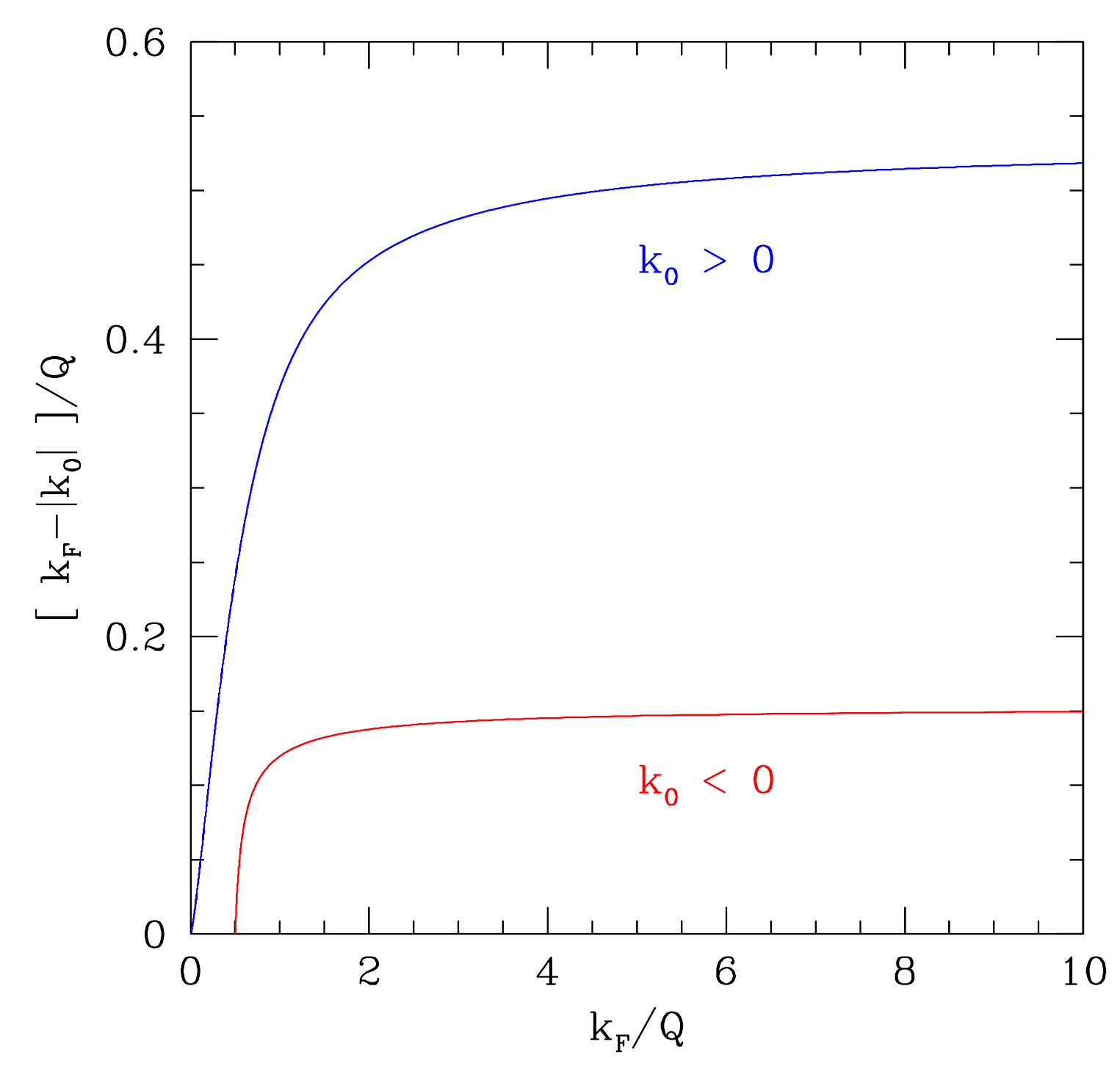}
\caption{Boundaries in parameter space where the steady state solution allows for a single black-hole horizon. The regions below either curve identify a stationary state characterized by a supersonic transition. The blue line corresponds to a left moving fluid ($k_0 >0$) while the red line to a right moving fluid ($k_0 <0$).}
\label{fig-asinto}
\end{center}
\end{figure}

The integrals can be easily evaluated for both the particle current and the asymptotic densities at $x\to\pm\infty$ and allow to evaluate the range of parameters leading to a steady state characterized by a single supersonic transition. Figure \ref{fig-asinto} summarizes the results: the blue curve refers to a left moving fluid ($k_0 >0$) with subsonic velocity at $x\to +\infty$ which gets accelerated by ``falling into the waterfall" and becomes supersonic at $x \to -\infty$ if $k_F-k_0$ lies below the blue line. Instead, the red curve shows that a right moving fluid ($k_0<0$), which starts subsonic at  $x \to -\infty$, can get supersonic flowing against the step potential if $k_F+k_0$ belongs to the region below the red line. In this case, however, $k_F$ must exceed the minimum value $\frac{Q}{2}$. In both cases, a black-hole horizon forms close to $x=0$, only for sufficiently high initial velocities: $|k_0|\lesssim k_F$.

\begin{figure} [ht!]
\begin{center}
\includegraphics[width=8cm]{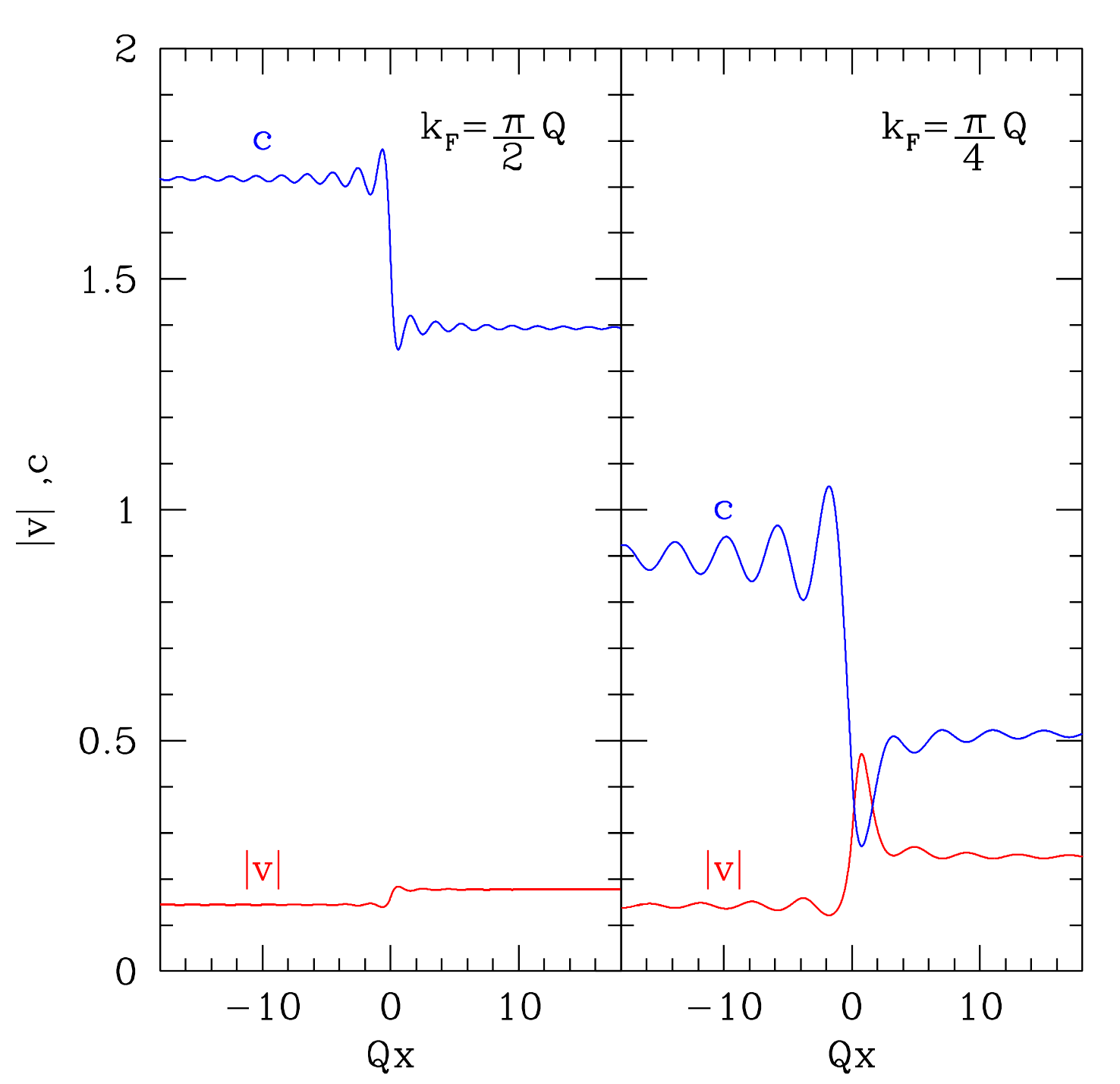}
\caption{Absolute value of the fluid velocity (red) and of the sound speed (blue) in the stationary states corresponding to $k_0=0$ and two different Fermi momenta: $k_F=\frac{\pi}{2}Q$ (left panel) and $k_F=\frac{\pi}{4}Q$ (right panel). Velocities are in unit of $\frac{\hbar Q}{m}$.}
\label{fig-0}
\end{center}
\end{figure}

The steady states corresponding to an initially still Fermi gas, i.e. a vanishing $k_0$, display a finite particle current due to the sudden appearance of the waterfall potential, leading to a leftwards motion (i.e. with negative velocity) of the fluid. For this case, two velocity profiles are shown in Figure \ref{fig-0} for $k_F=\frac{\pi}{2}Q$ and $k_F=\frac{\pi}{4}Q$: while in the former case the steady state is always subsonic, in the latter a supersonic flow sets in a finite region around $x\sim 0$, giving rise to a pair of black-hole/white-hole horizons, i.e., to a region where the flow passes from subsonic to supersonic and then becomes subsonic again.  

\begin{figure} [ht!]
\begin{center}
\includegraphics[width=8cm]{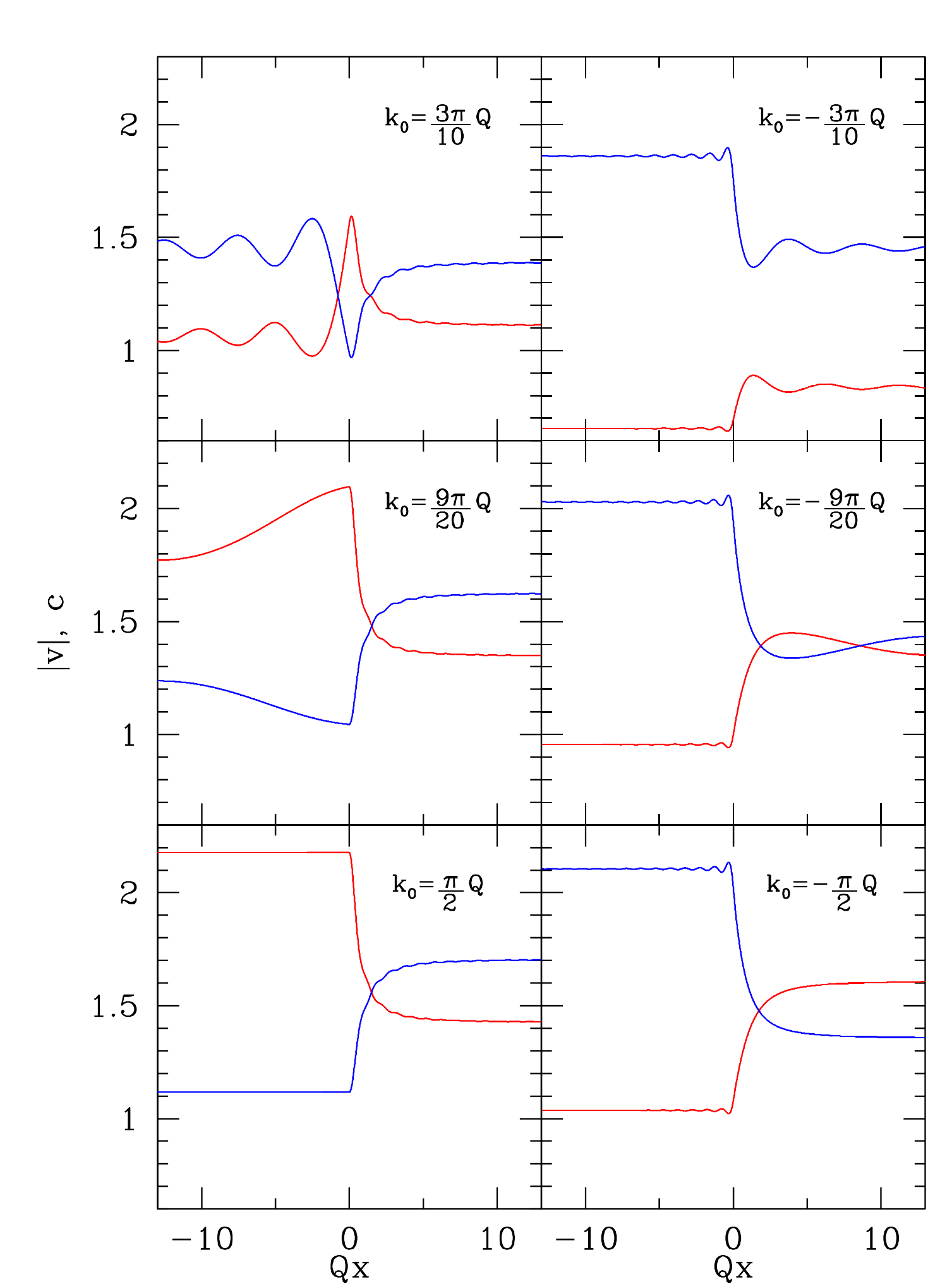}
\caption{Absolute value of the fluid velocity (red) and of the sound speed (blue) in the stationary states corresponding to $k_F=\frac{\pi}{2}Q$ and several values of $k_0$. Left panels correspond to a left moving fluid, right panels to right moving fluid (i.e. against the step). Velocities are in unit of $\frac{\hbar Q}{m}$.}
\label{fig-vel}
\end{center}
\end{figure}

When we start from a fluid which is already moving (i.e. $k_0\ne 0$), the long time characteristics of the flow velocity and of the sound speed show different possible behaviors. Illustrative plots of the stationary velocity profiles for a few representative choices of $k_0$ at fixed $k_F=\frac{\pi}{2}Q$ are shown in Figure \ref{fig-vel} for both cases of a fluid falling into the waterfall ($k_0 >0$, left panels) and flowing against the step potential ($k_0 <0$, right panels). While for $k_0=0$ Figure \ref{fig-0} shows that the fluid motion is subsonic, when $k_0$ is increased (i.e. when the fluid falls from the potential step) a small supersonic region appears close to the step (see the case $k_0=\frac{3\pi}{10}Q$). Further increasing $k_0$ (see $k_0=\frac{9\pi}{20}Q$) the flows becomes supersonic in the whole downstream (i.e. left) region. At $k_0=k_F=\frac{\pi}{2}Q$ the velocity becomes constant for all $x<0$, as previously discussed. The case of a fluid flowing against the step potential (right panels) is qualitatively similar, showing a fully subsonic flow for $k_0=-\frac{3\pi}{10}Q$, a black-hole/white-hole pair at $k_0=-\frac{9\pi}{20}Q$ and a single supersonic transition at $k_0=-k_F=-\frac{\pi}{2}Q$. In the latter case, however, the fluid velocity is not constant in the downstream (i.e. right) region. In all the cases, the undulations present both in the upstream and downstream regions are due to the presence of a sharp Fermi surface.

Note that the steady state solutions discussed here are not invariant by time reversal. Being the hamiltonian real, the complex conjugate of the single particle wavefunctions (\ref{asy}) are still exact eigenfunctions for the free Fermi gas; the time reversal of a {\sl scattering state}, however, is not a scattering state but is written as a particular linear combination of the two scattering states. Therefore, the Slater determinant of the time-reversed states describes a different stationary state of the system with the same density profile and opposite current $j$, as shown by Eqs. (\ref{rho},\ref{gei}): wherever the original solution displays a black-hole (BH) horizon, the time-reversed state predicts a white-hole (WH) horizon. However, we stress that, according to the analysis leading to Eq. (\ref{asy}), the time reversed state does not describe the stationary solution spontaneously reached by the system after a quench.

It is interesting to compare these exact results with the stationary states of the corresponding Gross-Pitaevskii equation,
to check whether the semiclassical approach is able to capture the variety of behaviors displayed by the TG gas model. 
Looking for solutions to  Eq. (\ref{gpe5}) of the form $\psi(x,t) =\Phi(x)\,e^{-\frac{i}{\hbar}\mu t}$ we obtain a 
non-linear differential equation whose solutions can be found analytically in the case of a step potential $V(x)=V_0 \,\Theta(x)$, where $\Theta(x)$ is the Heaviside function. The generic stationary solution depends on two parameters: the chemical potential $\mu$ and the current $j$. The equation has real coefficients and then each stationary solution has a time reversal partner with opposite current: the direction of the particle flow is reversed by keeping the same density profile. Restricting our analysis to solutions leading to density profiles asymptotically flat\footnote{A different class of solutions of the cubic GPE, characterized by an oscillating density profile, was discussed in Ref. \cite{pav}. Although this class of solutions is also present in the quintic GPE, we do not discuss their properties in detail.} for $x\to \pm\infty$, we generally find only monotonic decreasing density profiles and fully subsonic flows. Only for a special relationship between $\mu$ and $|j|$ a different solution appears, displaying a supersonic transition, which might represent either a BH or a WH horizon according to the sign of $j$. The density (and velocity) profile is flat for $x<0$ while is monotonically increasing for $x >0$: this solution corresponds to the ``half soliton" profile in full analogy with the known results valid for the standard (cubic) GPE \cite{pav}. The analytical form of the half soliton for the quintic GPE is $\Phi(x) = \sqrt{n(x)}\,e^{\frac{i}{2}\varphi(x)}$ where the modulus and the phase are expressed in terms of the unique parameter $\delta$ 

\begin{equation}
\delta^2 = 1 - \frac{j^2}{\pi^2 n_\infty^4} \, .
\end{equation}

Along the positive semi-axis $x>0$ we have 

\begin{eqnarray}
n(x) &=& n_\infty \,\left [ 1-\frac{3\,\delta^2}{2+\sqrt{4-3\,\delta^2}\,\cosh\left ( 2\pi\,\delta\,n_\infty x\right)} \right ] \, ,
\\
\varphi(x) &=& 2\pi\,\delta\,n_\infty\,x - 
\arcsin \left [ \frac{\alpha +\cosh\left ( 2\pi\,\delta\,n_\infty x\right)}{
\cosh\left ( 2\pi\,\delta\,n_\infty x\right)+\alpha} \right ] \, .
\end{eqnarray}

where $\alpha = (4-3\,\delta^2)^{\frac{1}{2}}\,(2-3\,\delta^2)^{-1}$ and the asymptotic density $n_\infty$ is expressed in terms of the parameter $\delta$ by the algebraic equation:

\begin{equation}
\left (\frac{2\pi\,n_\infty}{Q} \right )^2= \frac{18\,(1-\delta^2)}
{(4-3\,\delta^2)^2 + \sqrt{4-3\,\delta^2}\,(8-9\,\delta^2)} \, .
\end{equation}

\begin{figure} [ht!]
\begin{center}
\includegraphics[width=8cm]{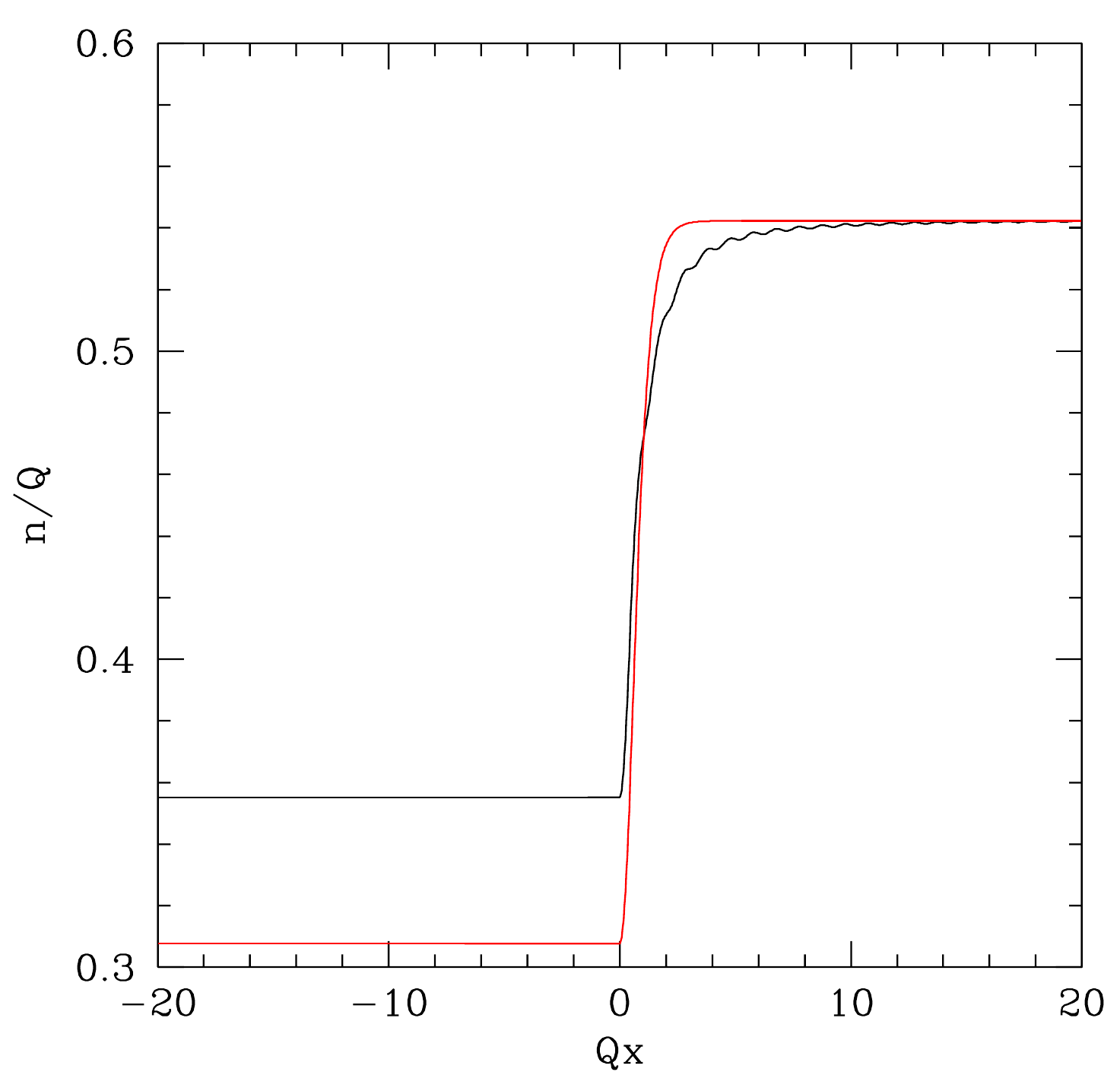}
\caption{Density profile for the exact stationary state at $k_F=k_0=\frac{\pi}{2}Q$ (black) compared to the half soliton solution of the quintic GPE corresponding to the same asymptotic density at $x \to +\infty$ (red).}
\label{fig-compare}
\end{center}
\end{figure}

For $x<0$, as previously stated, the density is constant and the phase linear in $x$. This analysis shows that the GPE is not able to reproduce, even qualitatively, the exact solution when a supersonic transition is present and the supersonic flows have non-monotonic profiles with a well defined limit for $x\to\pm\infty$ (see Figure \ref{fig-vel}). Only the special case $k_0=k_F$ appears to be related to the ``half soliton" solution of the GPE, although the agreement is not quantitative (see Figure \ref{fig-compare}): if the parameter $\delta$ is chosen so that the $x \to +\infty$ limit of $n(x)$ coincides with the exact value, the uniform density in the supersonic region is underestimated and the profile close to the transition point is not correctly reproduced. 

\section{Black hole formation}

We now examine the formation dynamics of an event horizon in this analogue model. We first note that all the steady state 
solutions we have found, together with their time reversal, are dynamically stable because the exact dynamics is linear and unitary, being that a free Fermi gas. A direct numerical integration of the Schr\"odinger equation starting from the analytical steady state solution confirms this expectation: a small random noise superimposed to the initial condition does not grow in time. This result might appear puzzling because we have seen that, for suitable choices of the parameters $(k_F,k_0)$ a black-hole/white-hole pair appears (see, for instance, the case $k_0=\frac{3\pi}{10}Q$ in Figure \ref{fig-vel}). Under these circumstances a dynamical instability is generally expected due to the black-hole laser mechanism \cite{ted,finazzi}, although detailed studies \cite{parentani1,parentani2,parentani3} found that additional requirements must be met in order to trigger the instability. The stability of the exact stationary states is shared by the half soliton solution of the GPE: a Bogoliubov-de Gennes analysis shows that, analogously to the known result valid for the dark soliton solution in the cubic GPE \cite{GPE,pav}, no complex eigenfrequency is present. The stability of the half soliton solution can be also easily checked numerically by direct integration of the GPE.

\subsection{Exact dynamics}

\begin{figure} [ht!]
\begin{center}
\includegraphics[width=8cm]{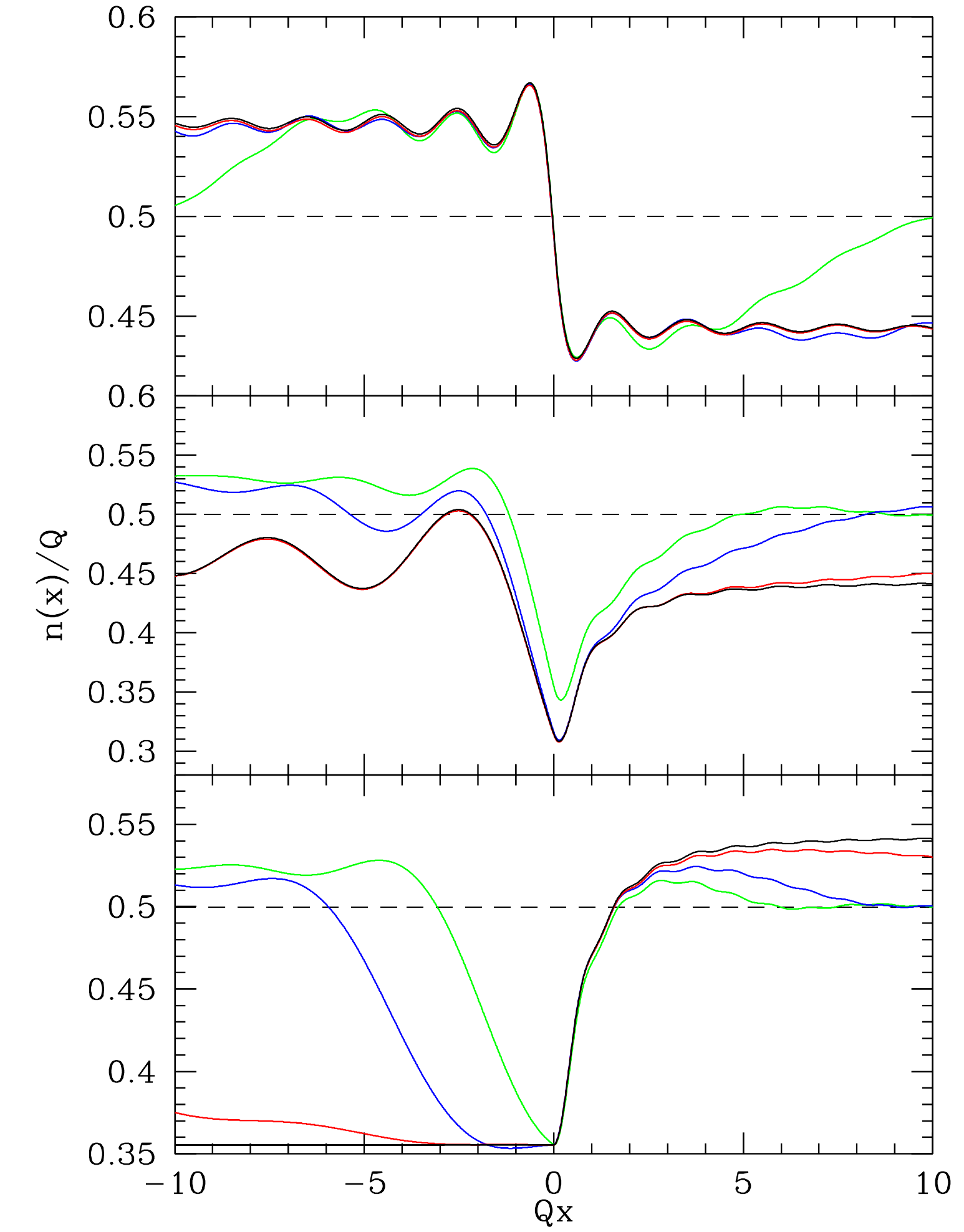}
\caption{Exact time evolution of the density profile for a left moving fluid with $k_F=\frac{\pi}{2}Q$. The dashed line shows the initial condition. Green, blue and red lines refer to evolution times $t=5,10,50$ respectively. The three plots correspond to different values of $k_0$: $k_0=0$ (top panel), $k_0=\frac{3\pi}{10}Q$ (central panel) and $k_0=k_F$ (bottom panel). The black line shows the steady states determined analytically according to the prescriptions of Section \ref{station}. The corresponding velocity profiles at stationarity are shown in Figure \ref{fig-vel} (left panels). The time unit is $\tau=\frac{m}{\hbar Q^2}$.}
\label{fig-fermi}
\end{center}
\end{figure}

Given the previous considerations, we might expect that a uniformly flowing TG gas perturbed by the sudden switching on of a step potential should evolve spontaneously towards one of the steady state solutions investigated in Section \ref{station}. Starting with a left moving fluid, this is indeed the case for the exact dynamics in all the different regimes previously identified as reported in Figure \ref{fig-fermi}, where a few snapshots of the evolving density profile are reported for $k_F=\frac{\pi}{2}Q$ and $k_0\ge 0$. 

As argued in our previous studies \cite{epl,prd} and shown in Figure \ref{fig-fermi}, the exact dynamics approaches the stationary state which we have determined by analytical means in the previous Section: i.e., the Slater determinant of the scattering states identified by the same parameters ($k_F,k_0$) of the initial condition. The time scale necessary to reach the steady state depends on the size of the region we are interested in; we see that a few tens of the natural time unit ($\tau=\frac{m}{\hbar Q^2}$) are enough to converge to the analytic solution in the range $|xQ|\lesssim 10$ and no shock waves appear during the exact time evolution. Thus, the dynamics drives a homogeneous, left moving TG gas to a known steady state which may exhibit the presence of sonic horizons, depending on how the initial density and velocity of the gas are chosen. Note again that the particular case of $k_0=k_F$ shows a constant density downstream and the half-soliton shape discussed in the previous Section.

\subsection{Semiclassical dynamics}

\begin{figure} [ht!]
\begin{center}
\includegraphics[width=8cm]{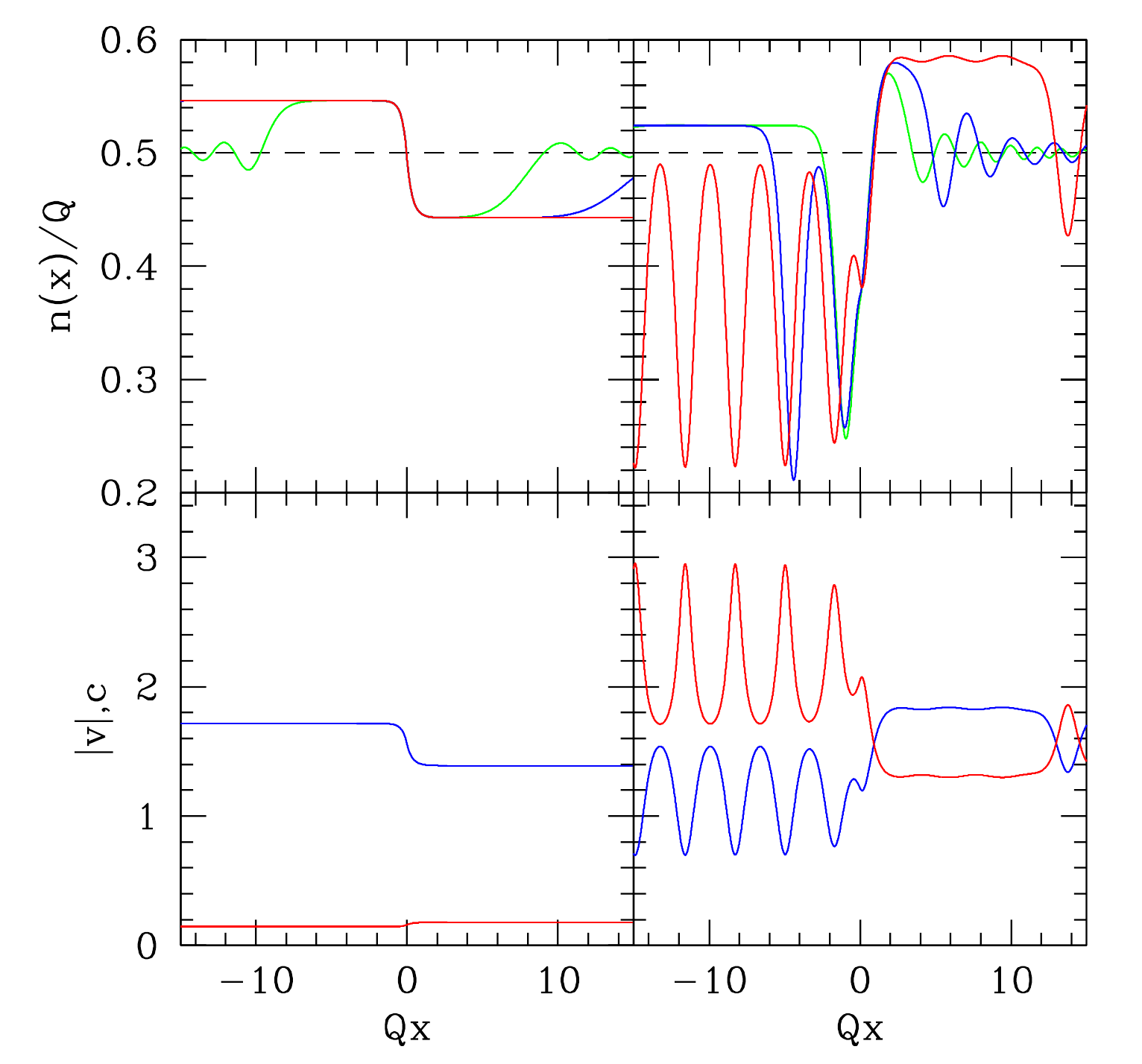}
\caption{Top panels: time evolution of the density profile according to the GPE dynamics for a left moving fluid with $k_F=\frac{\pi}{2}Q$. The dashed line shows the initial condition. Green, blue and red lines correspond to evolution times $t=2,10,50$ respectively for $k_0=0$ (left panel) and $k_0=k_F$ (right panel). The time unit is $\tau=\frac{m}{\hbar Q^2}$. Lower panels: fluid velocity (red) and sound speed (blue) at the latest time for $k_0=0$ (left) and $k_0=k_F$ (right). Velocities are in unit of $\frac{\hbar Q}{m}$.}
\label{fig-gpe}
\end{center}
\end{figure}

Remarkably, the case is quite different for the semiclassical dynamics, as shown in Figure \ref{fig-gpe}.

We know that only the fully subsonic stationary states and the half soliton solution have a direct analogue in the GPE, while there is no semiclassical solution corresponding to non monotonic density profiles and a well defined asymptotic limit at $|x|\to\infty$. For an initial condition leading to a subsonic steady state (see for instance the $k_0=0$ case), the GPE does indeed proceed smoothly, converging to a stationary state as expected. Furthermore, if we compare Figure \ref{fig-gpe} with Figure \ref{fig-0} and \ref{fig-fermi}, we see that the agreement between the semiclassical solution and the exact one is also quantitative, as the values of the density, of the fluid velocity and of the speed of sound show good agreement.

Instead, when the initial velocity increases, the semiclassical evolution does not approach a steady state even in the case $k_0=k_F$ (which is shown in Figure \ref{fig-gpe}), where we have shown that the half soliton solution is stable and closely mimics the exact result; here, instead, a sonic transition is formed during the evolution, as expected, but the density profile shows oscillations in space and time in the supersonic region and the flattening of the profile does not occur. Moreover, by carefully examining the dynamics, it can be noticed that the envelopes constituting this pattern are emitted one at the time, they have equal width, they move downstream with a constant velocity and their shape does not change during the evolution: they thus resemble the soliton trains present in the cubic GPE \cite{train1,train2}. This may not be evident from Figure \ref{fig-gpe}, where the downstream behavior looks more like a regular oscillation than a collection of solitons, but the structure becomes much clearer if we take lower values of $k_0$: in these cases the single envelopes are well separated and the succession of solitons is well defined. On the other hand, when we increase $k_0$ the interval between the emissions decreases but we recover the same qualitative behavior for all the intermediate cases $k_0 < k_F$ which feature a sonic horizon: there, in fact, the semiclassical equation does not allow for stationary solutions and indeed the soliton emission prevents to attain a steady state up to the $k_0=k_F$ limiting case, where a stationary solution of the GPE (the half soliton) is present but is not reached by the GPE dynamics.

\section{In search of the white hole}

A white hole is the time reversal of the black hole \cite{kruskal} and we already noted that the corresponding stationary state is indeed a stable solution of the exact dynamical equations, which is in agreement with previous studies on analogue white holes in BECs \cite{parentani4,mayoral}. So one might expect that for each initial condition leading to the formation of a black-hole horizon, starting from a time reversed initial state, the dynamical evolution of a free uniform TG gas would give rise to a white hole. In other words, if an initial condition defined by the two parameters $(k_F,k_0)$ leads to the formation of a black hole, does the time evolution of the state $(k_F,-k_0)$ give rise to a white-hole horizon? 

\subsection{Exact dynamics}

\begin{figure} [ht!]
\begin{center}
\includegraphics[width=8cm]{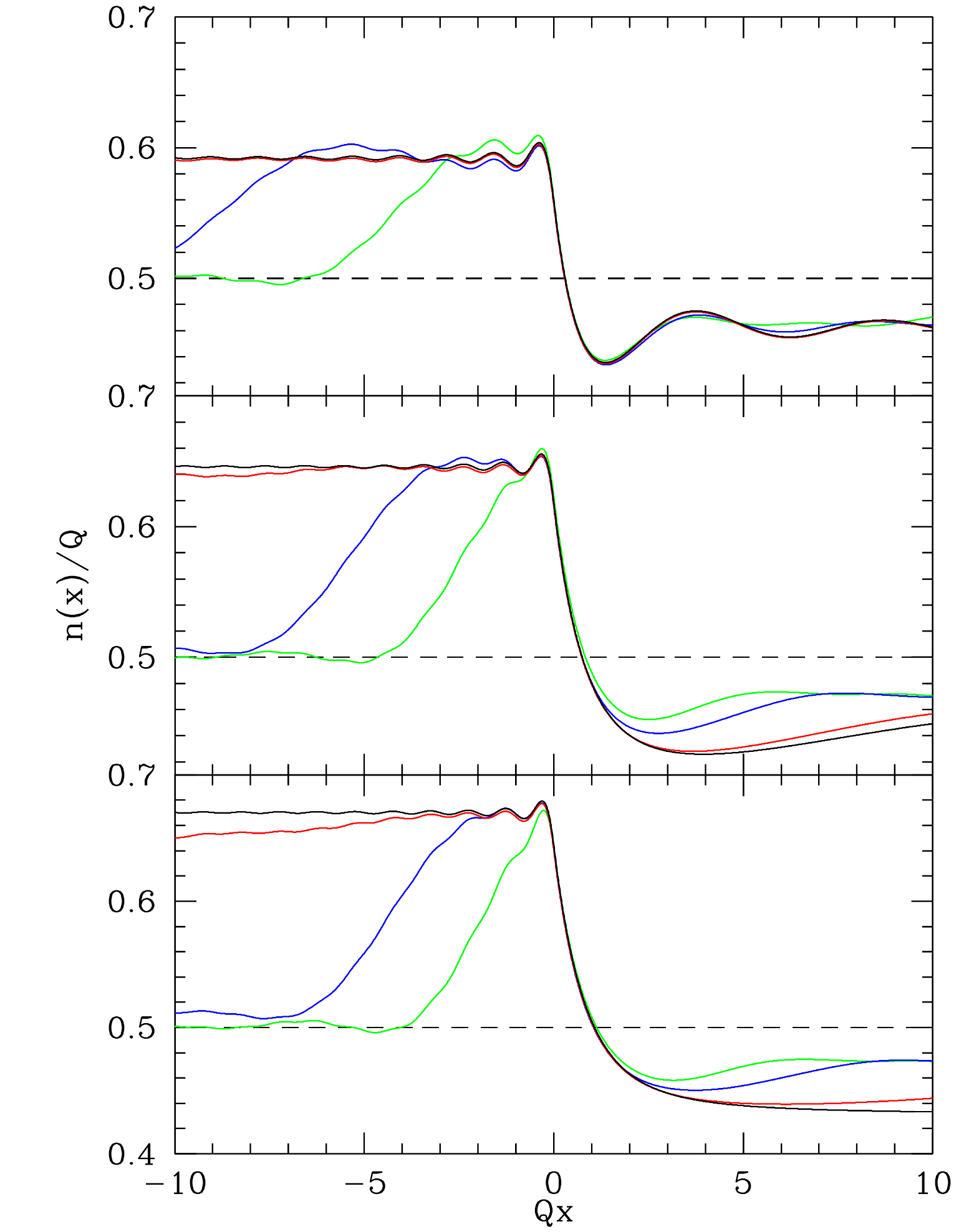}
\caption{Exact time evolution of the density profile for a right moving fluid with $k_F=\frac{\pi}{2}Q$. The dashed line shows the initial condition. Green, blue and red lines refer to evolution times $t=5,10,50$ respectively for $k_0=-\frac{3\pi}{10}Q$ (top panel), for $k_0=-\frac{9\pi}{20}Q$ (central panel) and for $k_0=-k_F$ (bottom panel). The black line shows the steady states determined analytically according to the prescriptions of Section \ref{station}. The corresponding velocity profiles at stationarity are shown in Figure \ref{fig-vel} (right panels). The time unit is $\tau=\frac{m}{\hbar Q^2}$.}
\label{fig-right}
\end{center}
\end{figure}

In order to test this claim, we have to investigate whether the dynamics drives a right moving fluid flowing against a step potential towards a white-hole configuration. This is not the case, as already shown in Figure \ref{fig-vel}, where the properties of the steady states identified by the pairs $(k_F,\pm k_0)$ are compared. To better illustrate this point, in Figure \ref{fig-right} we plot the exact dynamical evolution of a TG gas starting from different initial conditions defined by $k_F=\frac{\pi}{2}Q$ and $k_0<0$, i.e. a right moving uniform fluid.

When the potential step is switched on the density profile smoothly converges towards the analytical steady state shown in the right panels of Figure \ref{fig-vel}. The stationary flow may then be fully subsonic (see the case of $k_0=-\frac{3\pi}{10}Q$), may display a finite supersonic region (see $k_0=-\frac{9\pi}{20}Q$) or give rise to a black-hole horizon (see $k_0=-k_F=-\frac{\pi}{2}Q$). We never observe the spontaneous formation of a white-hole horizon, i.e. a flow going from supersonic to subsonic when passing through the potential step: the external potential always impresses an acceleration on the flow, irrespective of the direction of motion. The observed dynamics can be understood in terms of scattering states: a white hole is formed by taking the time reversal of a black hole, which, due to the dynamics induced by the step, is a particular superposition of different states. Thus, a initially homogeneous fluid (i.e. a single ``incoming'' state) can never lead to a white-hole configuration.

\subsection{Semiclassical dynamics}

\begin{figure} [ht!]
\begin{center}
\includegraphics[width=8cm]{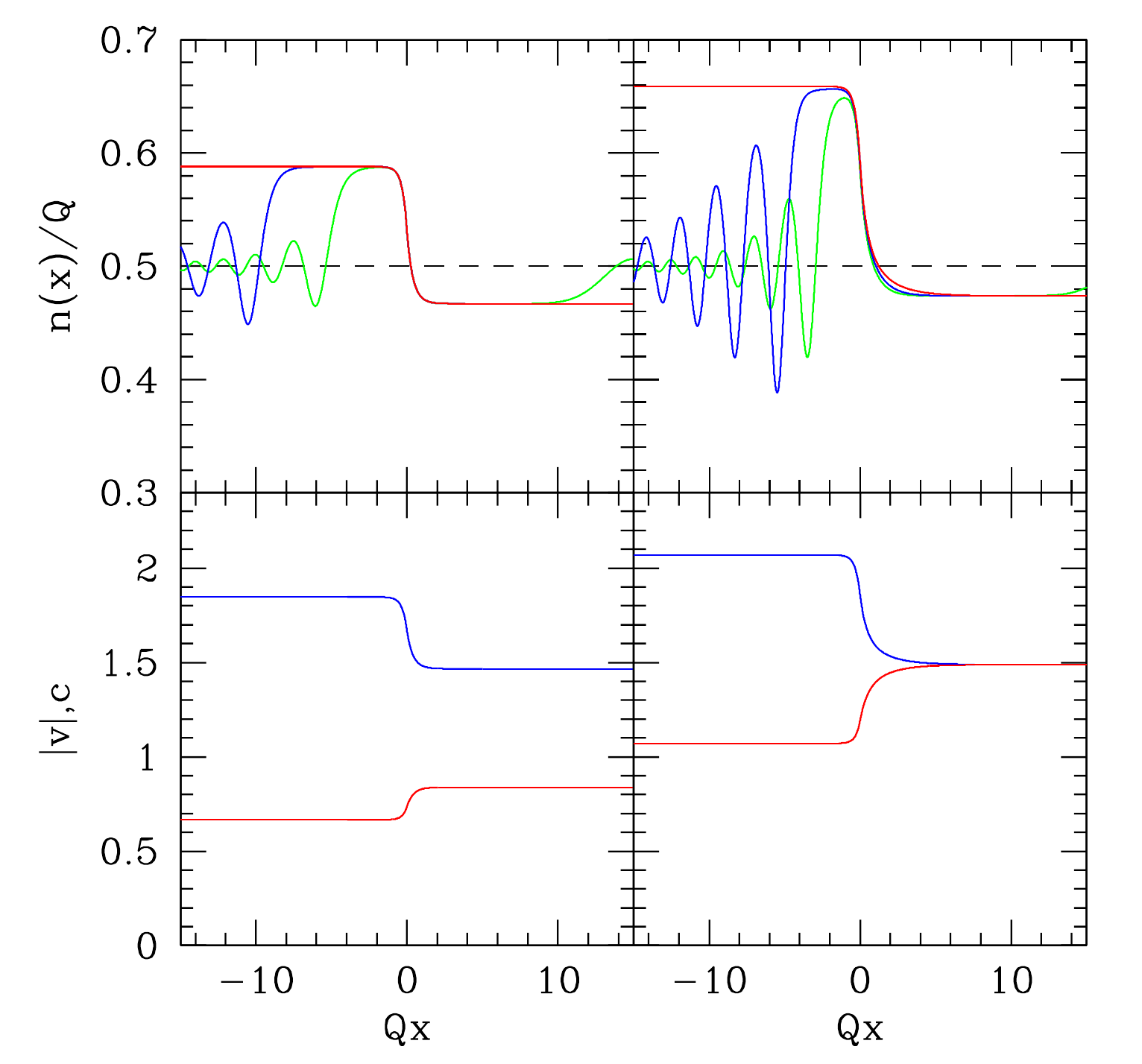}
\caption{Top panels: time evolution of the density profile according to the GPE dynamics for a right moving fluid with $k_F=\frac{\pi}{2}Q$. The dashed line shows the initial condition. Green, blue and red lines correspond to evolution times $t=5,10,50$ respectively for $k_0=-\frac{3\pi}{10}Q$ (left panel) and $k_0=-k_F$ (right panel). The time unit is $\tau=\frac{m}{\hbar Q^2}$. Lower panels: fluid  velocity (red) and sound speed (blue) at the latest time for $k_0=-\frac{3\pi}{10}Q$ (left) and $k_0=-k_F$ (right). Velocities are in unit of $\frac{\hbar Q}{m}$.}
\label{fig-gperight}
\end{center}
\end{figure}

Also in this case, the GPE dynamics of a right moving fluid differs significantly from the exact result. In Section \ref{station} we showed that, while the GPE admits a class of stable subsonic stationary solutions, only the ``half soliton" wavefunction displays a sonic horizon, which may correspond either to a BH or a WH horizon according to whether the mass current is negative or positive. Figure \ref{fig-gperight} shows the main features of the dynamics according to the GPE: when the initial condition corresponds to a generic velocity in the range $-k_F<k_0<0$, the evolution proceeds smoothly. Density waves are ejected towards $x\to -\infty$ and a subsonic stationary flow is reached at long times (see left panels). In the limiting case $k_0=-k_F$ (right panels) the stationary flow becomes asymptotically sonic, with $v=c$ at $x\to +\infty$. 

It is interesting to notice how, also in this case, the semiclassical picture agrees with the exact one also from a quantitative point of view only in the subsonic regimes: indeed, by comparing Figure \ref{fig-gperight} with Figure \ref{fig-vel} and \ref{fig-right} it is clear how the values of the density, the gas velocity and the sound speed are comparable whenever the flow is subsonic. Yet, when a supersonic transition sets in, the quantitative agreement is spoiled.\\

\begin{figure} [ht!]
\begin{center}
\includegraphics[width=8cm]{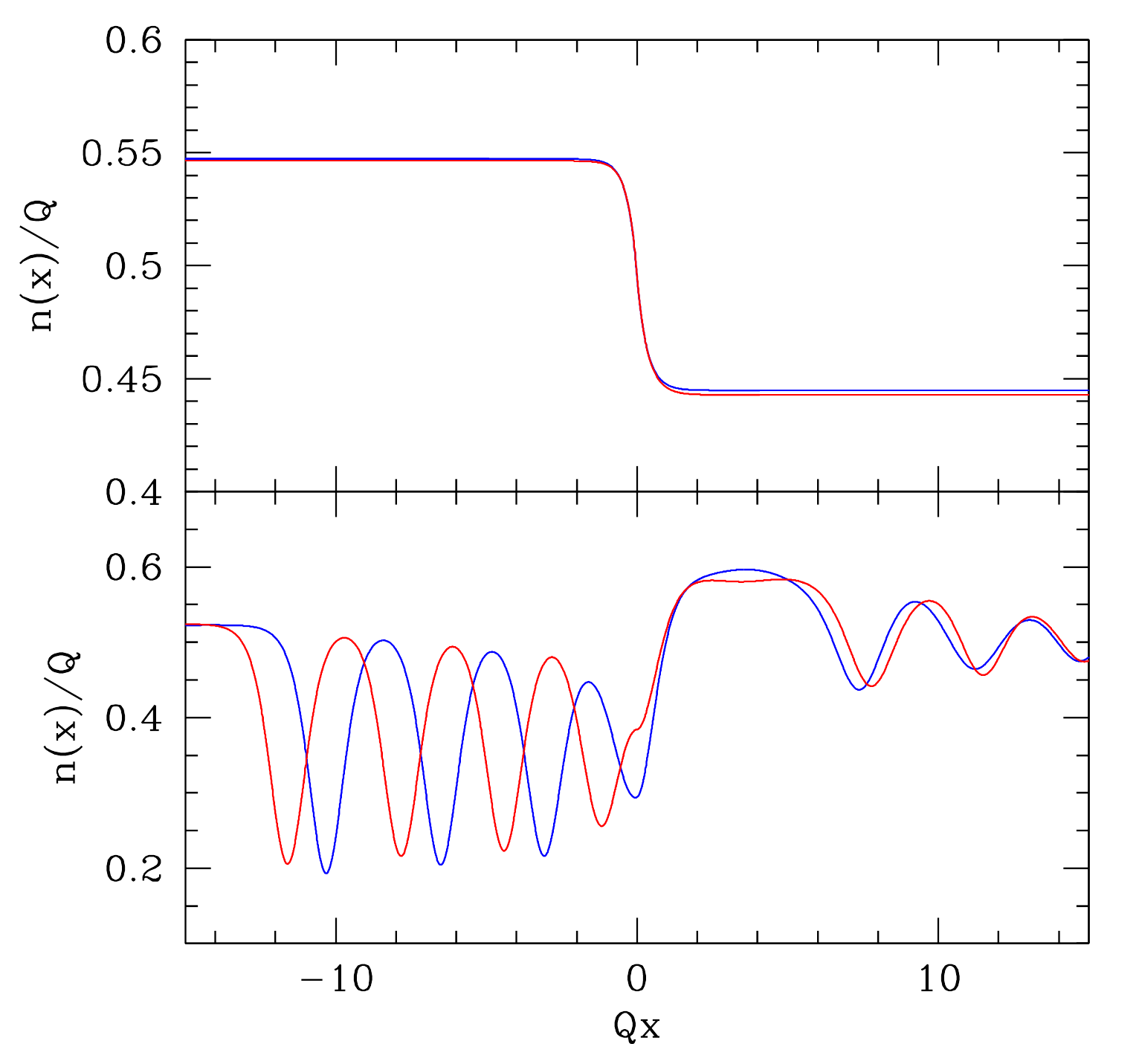}
\caption{The two figures compare the long time behavior of the density when the usual (cubic) GPE is used (blue curves) in contrast with the modified (quintic) form of Eq. (\ref{gpe5}) (red curves). In both cases, the same dimensionless coupling constant $\tilde g=\pi^2$ has been used. The profiles are qualitatively the same for both plots. The top plot shows a subsonic flow with $k_0=0$ and $k_F=\frac{\pi}{2}Q$; the bottom plot, instead, is the case in which a sonic horizon form, i.e. $k_0=k_F=\frac{\pi}{2}Q$. Similar results are obtained in the case of a right moving fluid. All the curves refer to a time $t=20\frac{m}{\hbar Q^2}$.}
\label{gpe35}
\end{center}
\end{figure}

Finally, we want to address an important point regarding the semiclassical treatment: recalling Eq. (\ref{gpe5}), the question may arise whether the dynamical behavior of the gas in presence of a sonic transition may be due to the form of the non-linearity, while an approach based on the canonical GPE would have given a different picture. Figure \ref{gpe35} shows that this is not the case, as the main features (horizon formations, oscillations etc.) observed in the semiclassical dynamics are not a peculiarity due to the precise form of Eq. (\ref{gpe5}). Indeed, the same qualitative behavior is retrieved when the above analysis is repeated with the standard (cubic) Gross-Pitaevskii equation and the quantitative differences between the two cases are irrelevant to the conclusions drawn in this study. This is important as theoretical studies on sonic horizons in BECs usually rely on the standard GPE and a description of the dynamical behavior of a condensate in these regimes has been deemed useful several times in this context \cite{pav,recati,larre}. To the best of our knowledge, only two other works investigated the formation dynamics of the sonic horizons \cite{parentani_latest,nohair} but the results only partially agree with ours due to the particular configurations chosen\footnote{In \cite{parentani_latest,nohair}, the gas is chosen with an interaction amplitude $g$ which varies along the axis of motion. Here, instead, we explore the case of a fixed $g$.}.

\section{Further possible configurations}

The analysis carried forward in the previous Sections starts from the assumption that $|k_0| \le k_F$, which means that initially, when the potential is switched off, the fluid has a subsonic (at most sonic) velocity. For completeness, we will now relax this hypothesis and investigate what happens if the gas initially flows at supersonic speeds.
Eqs. (\ref{analytic1},\ref{analytic2},\ref{analytic3}) provide the expected density profile and the particle current at stationarity also for $|k_0| > k_F$.

Following the previous discussion, we can study the dynamics of the gas after the quench numerically and we can investigate the possible formation of a sonic horizon both within the exact dynamics and according to the semiclassical (GPE) approximation. One would expect that a supersonic homogeneous gas subject to an external waterfall potential would not give rise to a sonic transition, but, instead, the switching-on of the step potential introduces various interesting configurations as we will now describe in some detail. 

\begin{figure} [ht!]
\begin{center}
\includegraphics[width=9cm]{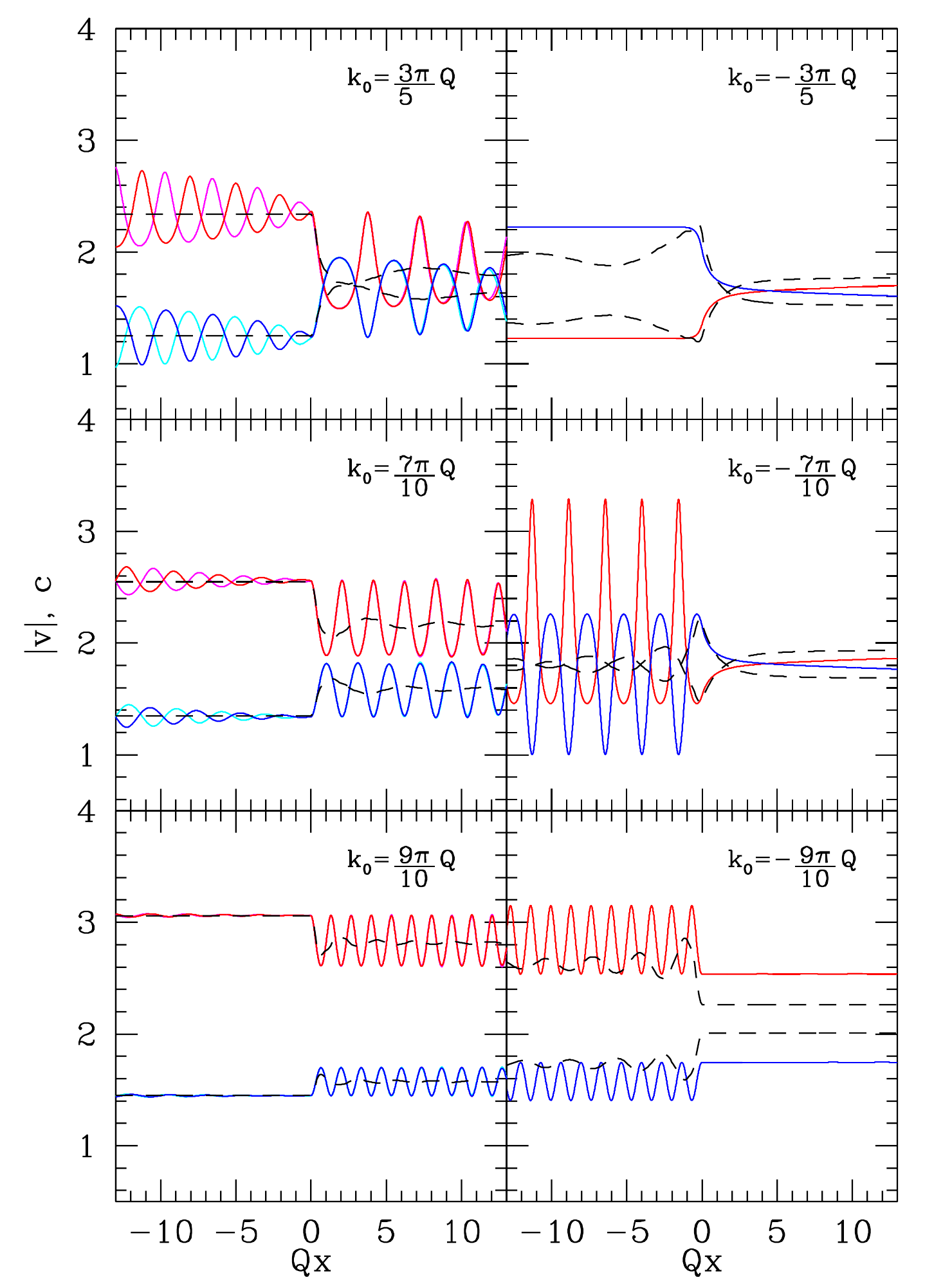}
\caption{Absolute value of the fluid velocity (red) and of the sound speed (blue) according to the GPE dynamics. The initial state is a homogeneous fluid with $k_F=\frac{\pi}{2}Q$ and different values of $k_0$. Left panels correspond to a left moving fluid ($k_0>0$), right panels to right moving fluid ($k_0<0$, i.e. against the step). Velocities are in unit of $\frac{\hbar Q}{m}$. The snapshots are taken at a time $t=60 \frac{m}{\hbar Q^2}$ for the right moving cases, where the GPE slowly drives the system towards a stationary state. For a left moving fluid, the snapshots correspond to $t=15 \frac{m}{\hbar Q^2}$. In these cases, the system reaches a steady state in the upstream region, while the velocity profiles oscillate in time (and space) downstream. Snapshots taken at a time $t=14 \frac{m}{\hbar Q^2}$ are also shown (magenta and cyan lines) to emphasize the temporal oscillations. The black dashed line represents the stationary state reached by the exact evolution in each case.} 
\label{supersonic-figure}
\end{center}
\end{figure}

Let us start with a left moving gas flowing over the potential step. If we choose a value of the initial gas velocity which is slightly bigger than the sound speed ($k_0 \gtrsim k_F$), the exact evolution drives the system towards a stationary state displaying a sonic horizon. By further increasing $k_0$ this region shrinks until the fluid is supersonic on the whole axis. The dashed line of Figure \ref{supersonic-figure} illustrates this behavior: indeed, the upper left panel shows a value of the initial velocity for which the subsonic region consists in one point, while in the central and bottom left panel the flow is always supersonic. Instead, the long time behavior of the semiclassical (GPE) solution shows an oscillation pattern in the upstream region and a train of solitons in the downstream region, thus never leading to the exact stationary state. This is found to be true for any value of $k_0 > k_F$, as Figure \ref{supersonic-figure} shows. Also, as noted in Section \ref{station}, under this circumstances the exact solution is always flat in the downstream region. \\
The right panels of Figure \ref{supersonic-figure}, instead, report the long time evolution of a right moving fluid. The exact results (dashed lines) show that in a range of velocities $|k_0| > k_F$ a BH horizon is formed: the flow upstream is subsonic and a supersonic transition occurs near the position of the step. By further increasing the value of the initial velocity the subsonic region becomes finite until it disappears, leaving a supersonic flow everywhere. Interestingly, it can be seen that both the exact and the semiclassical solution tend to a finite value in the downstream region while the two are significantly different in the upstream region: while the exact solution has smooth behavior, the semiclassical one develops spatial oscillations. Also, as already noted in Section \ref{station}, the exact solution is always flat in the downstream region if $-k_0 > Q+k_F$ (bottom panel).

\section{Conclusions}

We have studied, both analytically and numerically, an exactly solvable model describing a flowing one-dimensional Bose gas, in order to shed light on the physics underlying the analogue gravity paradigm. This model of hard-core bosons enables to follow the formation dynamics of black- and white-hole horizons after an external step potential is switched on: the analogue of a gravitational collapse event. Indeed, the fluid starts from a situation of constant density and velocity and, after a quench, it reaches another stationary configuration. From a curved spacetime point of view, this process is the exact analogue of Hawking famous \textit{gedankenexperiment} \cite{hawking2}. Accordingly, a phonon flux emerging from the black-hole horizon is expected \cite{epl,prd}. The presence of an outgoing heat flux in a stationary configuration is indeed possible in our one-dimensional geometry because the horizon cuts the space in two infinite regions. Furthermore, the shape of the external potential represents a realistic reproduction of existing experimental setups \cite{stein2010,stein14,stein2016,stein18,stein19} and a model used in various theoretical works (for example \cite{larre,recati,parentani_latest,nohair}).

We analytically found the stable stationary states of this model which allows for the existence of both black- and white-hole horizons, as well as black/white hole pairs. By numerical integration of the exact evolution equations we found that if a horizon is formed after the quench, either an isolated black-hole horizon or a tight black/white hole pair is formed but a white-hole horizon can never be reached dynamically. This somewhat conforms with the analogue gravity paradigm: the positive black-hole entropy allows for the spontaneous formation of an horizon, while the negative white-hole entropy does not. Moreover this study suggests that a sufficiently tight black/white hole pair is not necessarily unstable, at least in analogue models, while our solutions confirm that both a single black/white hole is a stable solution of the dynamical equations. 

Furthermore, we compared these findings with the solutions of a semiclassical equation usually adopted to describe this class
of models: i.e., the Gross-Pitaevskii equation, here suitably generalized to deal with a strongly interacting Bose gas. We found that only a subset of all possible stationary states are described by the GPE, including the subsonic solutions (with no event horizons) and the single black hole case, for a specific value of the mass current. However, the dynamics described by the GPE does not lead to the formation of a stationary solution whenever a sonic transition is present: indeed, starting from a uniformly flowing Bose gas, the semiclassical evolution gives rise to density waves propagating outwards which are not damped in time, preventing the convergence towards the stable stationary solution of the GPE. This behavior, which is a general feature also for the usual cubic GPE, suggests that results based on the integration of non-linear Schr\"odinger equations should be critically considered, at least as far as the long time dynamics is concerned.

Finally, to complete the description, we show other possible configurations leading to the formation of one (or more) sonic horizon which can be achieved by further increasing the initial velocity of the gas.

On a final note, the results obtained in this work provide an important step for future possible investigations as they give insights for new possible experimental configurations, they represent a key ingredient for the study of the hydrodynamical instability (and gravitational effect) known as black-hole laser and they help understanding the bounds on validity of semiclassical methods. To this extent, a question which arises from these results is what the optimal black-hole configuration should be in order to study the analogue of the Hawking effect. In fact, on one hand, as Hawking already pointed out \cite{hawking1}, the dynamical history of the gravitational collapse should not affect the emitted radiation, but, on the other hand, his line of thought heavily rests on the comparison between a precise initial and final stationary state, which we are able to identify through the exact solution of the model but not by means of semiclassical approaches.

\bibliographystyle{ieeetr}
\bibliography{Bibliography}

\end{document}